# On Privacy-Preserving Histograms

**Shuchi Chawla**[*]
Carnegie-Mellon

**Cynthia Dwork**
Microsoft Research

**Frank McSherry**
Microsoft Research

**Kunal Talwar**[†]
Microsoft Research

## Abstract

We advance the approach initiated by Chawla et al. for sanitizing (census) data so as to preserve the privacy of respondents while simultaneously extracting "useful" statistical information. First, we extend the scope of their techniques to a broad and rich class of distributions, specifically, mixtures of high-dimensional balls, spheres, Gaussians, and other "nice" distributions. Second, we randomize the histogram constructions to preserve spatial characteristics of the data, allowing us to approximate various quantities of interest, e. g., cost of the minimum spanning tree on the data, in a privacy-preserving fashion.

## 1 Introduction

In a census, individual respondents give private information to a trusted party (the census bureau), who publishes a sanitized version of the data. There are two fundamentally conflicting requirements: privacy for the respondents and utility of the sanitized data. Very roughly, the sanitization should permit the data analyst to identify strong stereotypes, while preserving the privacy of individuals.

A theoretical study of the census problem was initiated by Chawla et al. [4], who presented a definition of privacy that captures the compelling concept of *protection from being brought to the attention of others*. As the philosopher Ruth Gavison points out, not only is such protection inherently valuable, but its compromise invites further invasion of privacy, as every action of the target of attention can now be scrutinized. Thus, in [4] the goal of the adversary is to single out, or *isolate*, an individual. Very roughly, isolation occurs when an adversary produces a description that "looks" much more like one member of the database than it does any other. Conversely, if the adversary can only produce descriptions that simultaneously resemble many different members of the database, then intuitively the adversary has only constructed a coarse stereotype, and isolation has not occurred.

Histograms are ubiquitous in official statistics, and there is a vast literature on methods of perturbing contingency tables to ensure privacy. A central result in [4] states that if the data are drawn uniformly from a high-dimensional hypercube, then a simple technique, *recursive histogram sanitization*, preserves privacy: with overwhelming probability over the choice of the database, in the absence of auxiliary information[1], the probability that the adversary can isolate even one database point is exponentially small in the dimension. The probability space is over the random choices made by the sanitization algorithm and the adversary. The proof is quite robust and can tolerate many, but not all, kinds of auxiliary

---

[*] Research performed, in part, at Microsoft.
[†] Research performed, in part, at Berkeley.

[1] Information the adversary may have access to *other than* the sanitized database; e.g., complete or partial information for many of the points.

information[2].

We show that the approach proposed by Chawla et al. applies to an extremely broad and rich class of distributions, including Gaussians and uniform distributions on "nice" regions, such as cubes, balls, and spheres, as well as mixtures of such distributions. That is, we show that histograms of data sets drawn from mixtures of such "nice" distributions are useful, privacy-preserving, summaries of the data.

Given a histogram sanitization, not only can we learn counts, but we can also infer some information about where points are located. This spatial information can be exploited using known techniques to approximate certain features of the data set, such as the cost of a minimum spanning tree on the database points or the cost of an optimal solution to the facility location problem over the data points. The techniques require randomization of the histograms. Done naively this could damage the proof of privacy. Our second contribution is, therefore, to develop methods for randomizing the histogram constructions while provably maintaining privacy.

## 1.1 Summary of Results

**Review of Isolation and the Framework of [4].** A database is a collection of some number $n$ of (unlabeled) points drawn from $d$-dimensional space $\mathcal{R}^d$. We assume a distribution $\mathcal{D}$ on databases. The real database is chosen according to this distribution, and is denoted RDB.

A (possibly randomized) sanitization algorithm takes as input the RDB and produces as output a sanitized database, denoted SDB. The SDB may be of essentially any form, e. g., $n'$ points in some space (not necessarily the same space as the RDB), or a distribution over space. This framework is sufficiently general to allow sanitization via summaries, histograms, perturbation, etc.

The privacy adversary, called an *isolator*, has two inputs: the sanitized database SDB and auxiliary information $z$. The isolator may be randomized. On input (SDB, $z$) the isolator produces a point $q \in \mathcal{R}^d$. We can think of $q$ as a description of a candidate database point ($q$ need not actually be in the RDB; rather, it is the adversary's guess as to what someone in the RDB might look like). Intuitively, the isolator succeeds when $q$ looks much more like a particular real database point than it looks like any $t-1$ others. Here $t$ is a privacy parameter describing "many." A second parameter, $c$, formalizes the degree of resemblance.

**Notation.** We let $\|x - y\|$ denote the distance between two points $x$ and $y$ and $B(x, r)$ denote a $d$-dimensional ball of radius $r$ around point $x$.

**Definition 1.1.** [4] **(($c, t$)-isolation)** *Let $y$ be any RDB point, and let $\delta_y = \|q - y\|$. We say that $q$ ($c, t$)-isolates $y$ if $B(q, c\delta_y)$ contains fewer than $t$ points in the RDB, i.e., $|B(q, c\delta_y) \cap \text{RDB}| < t$.*

Chawla et al. [4] investigated the privacy of a recursive histogram sanitization procedure, in which the data lie in the $d$-dimensional hypercube of fixed side-length, say, 2, centered at the origin. The procedure takes the parameter $t \geq 2$ as input and is described as follows: cut the top-level hypercube into $2^d$ sub-cubes, of equal size, by splitting along the midpoint of each side; recurse on every sub-cube containing at least $2t$ points. This process results in a set of $d$-dimensional hypercubes of varying sizes. The sanitization is a description of the cuts made and the exact population of every resulting cell.

**Theorem 1.1.** [4] *Suppose that RDB consists of $n$ points drawn i.i.d. and uniformly from the cube $[-1, 1]^d$. There exists a constant $c$ such that the probability that an adversary, given a recursive histogram sanitization as described above, $c$-isolates even one RDB point is at most $2^{-\Omega(d)}$, independent of $n$[3].*

This is a powerful theorem; its weakness is that it applies to a single unrealistic distribution on data. This weakness is remedied here. We show the broad applicability of the approach of Chawla et al. by describing how to obtain histograms for well-rounded distributions. It is then straightfor-ward to obtain privacy-preserving histograms of mixtures of these distributions.

---

[2]Non-trivial sanitization against arbitrary auxiliary information is provably impossible for *any* non-trivial notion of privacy compromise) [5].

[3]Chawla et al. also gave a formal specification of what is required from a data sanitization algorithm: access to the sanitized data should not *increase* an adversary's ability to isolate any individual beyond what it can do given only the auxiliary information. In the absence of auxiliary information the theorem above suffices.

In the proof of the theorem, volume is used as a surrogate for uncertainty: the adversary is given information about the number of database points in a cell, but from the adversary's perspective the location of these points within a cell is uniform. The proof relies on a lemma stating that for appropriate values of $r$, the volume of the intersection of a ball $B(q,r)$ with a cell in the histogram grows exponentially with $r$ (the upper bound on $r$ depends on the diameter of the cell in question). For this reason we need histogram cells to be "well-rounded" (think of a ball and a strand of spaghetti: increasing the radius of the ball gives only a linear growth in the volume of the intersection with the spaghetti strand).

**Histograms for Mixtures of "Nice" Distributions.** Our constructions replace the subcubes-based subdivision used for hypercubes in [4] by an appropriate algorithm to subdivide "round" cells into smaller round cells. Chawla et al. prove that if for all points $q$, all radii $r$, and all cells $C$ in the histogram, one of the following conditions holds, then the probability that the adversary $(c,t)$-isolates any point in RDB is at most $\epsilon$. Here $P(C)$ denotes the parent cell of $C$.

$$B(q, cr) \supseteq P(C) \quad \text{or} \quad \frac{\text{Vol}(B(q,r) \cap C)}{\text{Vol}(B(q,cr) \cap C)} < \epsilon \quad (1)$$

In Section 2, we prove that for the subdivision algorithms described next, Condition 1 holds with a constant value of $c$ and $\epsilon = 2^{-\Theta(d)}$ (see Corollary 2.2 and Section 2.3). We now describe the subdivision techniques.

In the first method, we use a deterministic decomposition technique based on *nets*. Roughly speaking, we choose "centers" in succession from the cell so that they are *well-spread* (the distance between any two centers is not "too small") and the cell is tightly covered (the distance from any point in the cell to its nearest center is not "too large"). The subcells are then given by the Voronoi partition of the cell created by the centers. We argue that the Voronoi regions are relatively well-rounded, and so this technique gives nice privacy constants; however, the constants deteriorate with the depth of recursion, intuitively, because the cells become increasingly less well-rounded.

In the second method, we again subdivide cells by picking a set of centers and constructing a Voronoi diagram over these. However in this case, a carefully chosen number of centers is picked uniformly at random from the cell. The randomization allows us to obtain a good embedding of the dataset into a metric defined by the histogram, that protects privacy as well as preserves distances between points to a reasonable accuracy (as described in the following section). However, the bound on the privacy parameter $c$ in this case, is worse than that for the previous construction.

A third method, omitted for lack of space, embeds the $d$-dimensional ball into the $d$-dimensional hypercube, and then applies the histogram sanitization from [4] to the cube. In this case, recursion is not a problem, but the privacy parameter $c$ becomes $\Omega(d^2)$.

As in [4], our techniques can handle limited kinds of auxiliary information: even if the adversary has complete knowledge of a subset of the points, she cannot breach the privacy of the remaining points with probability larger than $2^{-\Theta(d)}$.

**Randomized Histograms.** The deterministic nature of previous histogram work allows for bad examples in which we are unable to compute fairly simple quantities accurately. For example, a simple arrangement of $2^d$ points near the center of the cube can ensure that in one step each sample is placed in its own cell and recursion terminates, reporting one element in each cell. While we might have liked to learn the diameter of the data set, or the cost of a minimum spanning tree, the adversarial arrangement prevents approximation to within any factor.

As has been noted elsewhere, many quantities of interest about a data set can be accurately approximated using *randomized* hierarchical subdivisions. The randomization prevents the adversarial arrangements and can give strong guarantees about approximate solutions.

We give two examples of randomized histogram constructions in Section 3, one using nested cubes for data drawn uniformly on the unit cube, and one based on nested spherical regions intended for data that is drawn from spherical distributions.

We will see in Section 3 that, much as randomization enables the preservation of distances in previous work on randomized embeddings [2], randomization will ensure a strong connection between the quality of approximations of certain quantities of interest and the so-called $t$-radii (distances to $t$-th nearest neighbors) of the database points. This matches our intuition: our notion of privacy demands uncertainty of each database point propotional to its $t$-radius. We don't *want* a solution that provides an adversary with less uncertainty than this.

### 1.2 Related Work

There is a vast literature on statistical disclosure control and data sanitization. An excellent survey is [1]. [4] contains a brief discussion of statistical techniques such as suppression, aggregation and perturbation of contingency tables, input perturbation in the statistics and data mining literatures, imputation, $k$-anonymity [13], and cryptographic approaches to privacy. There is also mention of interactive solutions, such as query auditing and output perturbation ([6], and, the more careful modern treatment [7, 3]). To our knowledge, the only work on (non-interactive) data sanitization, other than [4], that explicitly takes into account auxiliary information is the lovely paper of Efvimievski et al. [8].

There is a strong algorithmic literature on the power of probabilistic embeddings (see e.g. [2, 10, 11]). Applications include approximation algorithms, online algorithms and geometric data stream algorithms. We refer the reader to [11, 12] for a survey of some of these applications.

## 2 Histogram Sanitizations for Round Distributions

In this section we describe how to obtain histograms for round distributions, specifically, for well-rounded regions. Assuming the sanitizer knows the constituent distributions, it is straightforward to obtain privacy-preserving histograms of mixtures by sanitizing each region independently.

We define a standard, parameterized, notion of well-roundedness for regions. Broadly, our approach to sanitizing a well-rounded region will be to define a (recursive) histogram by choosing a set of *centers* (points in the region, not database points) and then subdividing the region according to the Voronoi partition induced by these centers.

We first argue that well-roundedness of the resulting cells is sufficient for the privacy proof of [4]; in particular, that Inequality 1 will be satisfied (Section 2.1). In Section 2.2 we describe conditions under which the Voronoi subdivisions remain (sufficiently) well-rounded after recursion. Finally, Section 2.3 describes two methods for choosing centers that satisfy the conditions of Section 2.2.

### 2.1 Privacy for well-rounded cells

We start with a definition for well-rounded cells.
**Definition 2.1.** *A cell $C$ is said to be $k$-well-rounded of radius $R$ iff $C$ is convex and $\exists p : B(p, \frac{R}{k}) \subseteq C \subseteq B(p, R)$.*

We now present the key lemma of this section that relates well-roundedness to the expansion property required for privacy.
**Lemma 2.1.** *Let $C$ be a $k$-well-rounded cell, $k \geq 1$, of radius $R$, and let $c' = \max\{2k, 2\sqrt{2}\}$. Then, for any point $q \in C$ and radius $r < \frac{R}{kc'}$, the following holds:*

$$\frac{Vol(B(q,r) \cap C)}{Vol(B(q,c'r) \cap C)} < 2^{-(\frac{d}{2}-1)}d = 2^{-\Theta(d)}$$

*Proof.* Let $p$ be as in the definition of well-roundedness. First consider the case when $q$ lies in the ball $B(p, \frac{R}{k})$. In this case, we will show that the volume of $B(q, c'r) \cap B(p, \frac{R}{k})$ is large. In particular this quantity is larger than the volume of a cap[4] of $B(q, c'r)$ that subtends an angle of $\delta = \pi/3$ at the center $q$, because $c'r < R/k$.

This volume can be computed as an integral over the volume of disks of thickness $c'r\theta$ and diameter $c'r \sin \theta$, with $\theta$ ranging from 0 to $\delta$. Some calculation shows that this volume is at least $\frac{\sin^d \delta}{\sqrt{\pi d}} Vol(B(q, c'r)) > \frac{2^{-d}}{2d} Vol(B(q, c'r))$. This implies that $Vol(B(q, c'r) \cap C) \geq Vol(B(q, c'r) \cap B(p, \frac{R}{k})) > \frac{2^{-d}}{2d} Vol(B(q, c'r))$.

---
[4]A cap of a $d$-dimensional ball subtending an angle $\delta$ at the center of the ball is defined by an axis through the center, and contains all points on the surface of the ball that subtend an angle at most $\delta$ with the axis. The volume of a cap is the volume of its $d$-dimensional convex hull.

Likewise, $\text{Vol}(B(q,r) \cap C) \leq \text{Vol}(B(q,r)) \leq (2\sqrt{2})^{-d}\text{Vol}(B(q,c'r))$, as $c' \geq 2\sqrt{2}$. Combining the two expressions, we get that for $q \in B(p, \frac{R}{k})$,

$$\frac{\text{Vol}(B(q,r) \cap C)}{\text{Vol}(B(q,c'r) \cap C)} < 2^{-(\frac{d}{2}-1)d}$$

Next consider the case when $q$ lies outside the ball $\mathcal{B} = B(p, \frac{R}{k})$. Consider the convex hull $H_q$ of $q$ and the ball $\mathcal{B}$. This lies entirely inside $C$, as $C$ is convex. Furthermore, apart from the ball $\mathcal{B}$, the convex hull contains a cone $\Lambda_q$ formed by tangents from $q$ to $\mathcal{B}$. Note that this cone subtends a large solid angle at $q$. In particular, the angle between any tangent and the line joining $q$ and $p$ is at least $\theta = \sin^{-1}\frac{1}{k}$ (by the well-roundedness of $C$).

Now we can compute the volume $\text{Vol}(B(q,c'r) \cap H_q)$ as the union of two terms—the intersection with $\Lambda_q$, and the intersection with $\mathcal{B}$. Note that when $q$ is far from $p$ (in particular, farther than $\frac{\sqrt{2}R}{k}$), then the intersection of $B(q,c'r)$ and $H_q$ is contained entirely inside the open-ended cone defined by $\Lambda_q$. This intersection has volume larger than the volume of a cap of $B(q,c'r)$ that subtends an angle $\theta$ at $q$, which is at least $\frac{1}{\sqrt{\pi}dk^d}\text{Vol}(B(q,c'r))$. On the other hand, when $q$ is closer than $\frac{\sqrt{2}R}{k}$ to $p$, the intersection of $B(q,c'r)$ and $\mathcal{B}$ has volume at least $\frac{1}{\sqrt{\pi}d2^{d/2}}\text{Vol}(B(q,c'r)) > \frac{1}{\sqrt{\pi}d(\sqrt{2}k)^d}\text{Vol}(B(q,c'r))$. Therefore, we get that $\text{Vol}(B(q,c'r) \cap H_q) > \frac{1}{\sqrt{\pi}d(\sqrt{2}k)^d}\text{Vol}(B(q,c'r))$.

Putting everything together, we get

$$\frac{\text{Vol}(B(q,r) \cap C)}{\text{Vol}(B(q,c'r) \cap C)} < $$
$$\frac{(2k)^{-d}\text{Vol}(B(q,c'r))}{(\sqrt{\pi}d(\sqrt{2}k)^d)^{-1}\text{Vol}(B(q,c'r))} < 2^{-(\frac{d}{2}-1)d}$$

□

The following corollary follows from the lemma by observing that for a point $q$ outside $C$, the ratio of volumes is bounded above by the corresponding ratio for an appropriate point $q'$ in $C$, albeit with $(c-1)$ in the place of $c$.

**Corollary 2.2.** *Let $C$ be a $k$-well-rounded cell of radius $R$. Then for any point $q$ and radius $r$, and for $c = 4k^2$, either $cr \geq R$, or,*

$$\frac{\text{Vol}(B(q,r) \cap C)}{\text{Vol}(B(q,cr) \cap C)} < 2^{-(\frac{d}{2}-1)d}$$

### 2.2 Voronoi-based histograms

Let $S$ be the region of interest, say, the $d$-dimensional unit ball or sphere. We start with the set $S$ as the level 0 cell in the histogram. At step $\ell = 1, 2, \ldots$, we consider all level $\ell - 1$ cells $C$ that contain more than $t$ points. For each of these cells, we obtain level $\ell$ cells by subdividing the cell as follows: we pick a set of centers in the cell, and construct a Voronoi partition of the space induced by these centers; this defines the next-level cells. Let $k_\ell$ be defined so that cells created at level $\ell$ are $k_\ell$-rounded. Different techniques for choosing the centers yield different values for $k_\ell$. In each of these, $k_\ell$ is an increasing function of $\ell$ (and not the number of points or cells).

We continue the recursion for a constant number of steps, or until all the cells have fewer than $t$ points, and release the exact counts of points in each cell. Assume the procedure runs for $s$ steps. Then each cell in this histogram is $k_s$-well-rounded for a constant $k_s$. Therefore, using the argument in the previous sections, privacy is achieved for a constant $c = 4k_s^2$.

**Definition 2.2.** *A set of points $\{p_1, \cdots, p_m\}$ is said to $r_1$-cover a set $C$, if $C \subset \cup_i B(p_i, r_1)$. It is said to be $r_2$-well-spread if for every pair of points $p_i$ and $p_j$, $i \neq j$, $\|p_i - p_j\| \geq r_2$.*

The following lemma gives conditions under which Voronoi cells are well-rounded (proof omitted for lack of space).

**Lemma 2.3.** *Let $C$ be a $k$-well-rounded cell of radius $R$, and $\{p_1, \cdots, p_m\} \subset C$ be a set of $r_2$-well-spread points that $r_1$-cover $C$ with $r_2 \leq r_1 < R$. Let $V_i$ be the region containing $p_i$ in the Voronoi partition of the points $\{p_1, \cdots, p_m\}$. Then the subcells $V_i \cap C$ are $\left(\frac{4r_1 k}{r_2}\right)$-well-rounded with radius $2r_1$.*

### 2.3 Two Methods for Choosing Centers

We now describe two methods for picking centers in cells so as to achieve the well-spread and covering properties. This along with the results in the previous section implies that the Voronoi-based histograms constructed from these centers preserve privacy.

**Method (1): Picking Well-Spread Centers Directly.** We describe the procedure as applied to a cell $C$. We pick points $p_1, p_2, \cdots$ in $C$ in succession as follows: the point $p_i$ is picked arbitrarily such that it is at distance at least $R/4$ from each of the points $p_j$ for $j < i$. We stop when every point in $C$ is within distance $R/4$ of at least one of the points $p_i$. By construction, this gives $R/4$-well-spread centers that $R/4$-cover the cell $C$. Then, at level $\ell$ of the recursion, this method gives us $k_\ell = 4^\ell k$, where $k$ is the roundedness-coefficient of the original set.

We do not know how to randomize this construction so as to be able to prove that the probability of a cell boundary lying between two real database points is proportional to the distance between the points (necessary in order to apply the results of the next section). This is addressed in Method (2), although the parameter $k$ for well-roundedness deteriorates more quickly with recursion than the corresponding one for Method (1).

**Method (2): Picking Centers Uniformly at Random.** As described earlier, in the second method, we pick centers from the cell at random. The following lemma shows that we obtain well-rounded Voronoi regions with a high probability.

**Lemma 2.4.** *Let $C$ be a $k$-well-rounded cell of radius $R$, and let $p_1, \cdots, p_m$ be $m = 4d8^d$ points picked uniformly at random from $C$. Then with probability at least $1 - exp(-d)$, the points are $\Omega(R/k)$-well-spread, and $R/4$-cover $C$. Consequently, the subcells $V_i \cap C$ are $O(k^2)$-well-rounded with radius $R/2$.*

Using this method, we get a well-roundedness-coefficient of $k_\ell = k^{2^\ell}$ at level $\ell$ of the recursion, where $k$ is the roundedness-coefficient of the original set.

The principal open question raised in this paper is whether it is possible to improve the techniques *or the analysis* described above to arrest the decay in the privay parameter at deeper levels of recursion.

## 3 Randomized Histograms

Given a histogram sanitization, not only can we learn counts, but we can also infer some information about where points are located. This spatial information can be used to approximate certain quantities of interest, such as the cost of a minimum spanning tree on the database points.

We have already noted in the Introduction that it is easy to construct examples in which the inferences drawn from the histogram sanitization are grossly pessimistic, and that randomization can be used to overcome this. If done carelessly, this will yield poorly-rounded cells, hurting the proof of privacy. In this section we show how to randomize the histogram construction *while preserving privacy*. We actually obtain results with additive, rather than the more traditional multiplicative, error.

The histogram distance $d_H(x, y)$ between points $x$ and $y$, taken to be the maximum distance between their two smallest containing cells, is fairly easily bounded using the diameters of these cells. Letting $\Delta_x$ and $\Delta_y$ be the diameters of the smallest histogram cells containing $x$ and $y$, respectively, the triangle inequality tells us that

$$\|x - y\| \leq d_H(x, y) \leq \|x - y\| + \Delta_x + \Delta_y \quad (2)$$

We now consider two classes of randomized histograms, and give bounds on the expected distance between any two histogram sanitized points, via bounds on $E[\Delta_x]$ in terms of the $t$-radius of $x$. As argued in [2, 10], expected distance bounds give useful bounds on any optimization problems whose output is a linear function of interpoint distances. Problems such as minimum spanning tree, facility location, and minimum weight matching fall into this framework.

To see why expectation bounds suffice, consider the relationship between an optimal solution on the actual distances and an optimal solution when using histogram distances. The actual cost of the histogram optimal solution is at most its histogram cost, and as this solution is optimal for the histogram distances this is in turn at most the histogram cost of the actual optimal solution. Using our bounds on expected distance increase (Equation 2), the histogram cost of the actual optimal solution is in expectation not much more than its actual cost. Therefore, the difference between the actual cost of the histogram solution and the optimal cost is small in expectation.

## 3.1 Randomized Histograms on the Unit Hypercube

We now look at a modification of the recursive histogram approach of Chawla et al. [4], who construct histograms using recursive subdivision of hypercubes. Recall that this technique subdivides any hypercube that contained at least $2t$ samples into its $2^d$ constituent hypercubes of half the edge length; this process continues until each cube contains fewer than $2t$ samples, at which point the counts of samples in each cube are released.

To randomize this construction and yield bounds on the expected increase in distances, we start from a construction that has been previously studied in the metric embeddings literature. The main idea is to conduct a standard hypercube subdivision, but on an inflated hypercube centered at a randomly chosen point inside the original hypercube. The side length of the inflated cube is twice that of the original. This ensures that the inflated cube completely covers the original. We can think of this as as covering space with a mesh, say, of edge-length 1, at an arbitrary offset from the origin. The subdivision is refined by using a smaller mesh, say, of edge-length $1/2$, *without changing the offset from the origin*. Only regions containing at least $2t$ points are refined.

As we will see in Theorem 3.1, this random translation will provide us with utility, but it may lack privacy. Specifically, for any cell in the histogram the points are distributed uniformly over the intersection of the cell and the unit hypercube. This is not a problem for interior cells where the intersection is the cell itself, a shapely hypercube, but those cells that intersect the surface of the hypercube may have very large aspect ratios leading to distributions that have low effective dimension and are easy to attack. To correct this, for every level of the histogram, all cells intersecting the hypercube surface are disbanded, and the interior cells exposed are expanded to cover the area of any adjacent discarded cells. The resulting subdivision into hyperrectangular regions has guaranteed aspect ratios, accomodating the privacy proofs in [4].

As noted previously, bounds on the expectation of $\Delta_x$ and $\Delta_y$ suffice, which we now provide. Let $r_x^t$ denote the $t$-radius of a point $x$: $r_x^t = \mathrm{argmin}_r\{|B(x,r) \leq t|\}$.

**Theorem 3.1.** *For any sample $x$, letting $\Delta_x$ be the diameter of the smallest cell containing $x$,*

$$E[\Delta_x] \leq 2\min\{d^{\frac{3}{2}}, td\} r_x^t \log(1/r_x^t) .$$

*Proof.* The diameter of the smallest cell is, up to a factor of 2, dominated by the diameter of the smallest cell in the traditional randomized histogram construction, where we do not collapse perimeter cells. We therefore shift our attention to that construction. We consider the contribution to the expectation from each level of the histogram. Then the total expectation is their sum. At level $i$, we can upper bound the probability that the recursion will terminate from lack of neighbors by the probability that the neighborhood $B(x, r_x^t)$ of $x$ is cut by the decomposition. To bound this, notice that it is at most the sum of the probabilities that it happens in each of the dimensions. In each dimension, the probability that $B(x, r_x^t)$ is cut is simply the length of the projection of $B(x, r_x^t)$ along that dimension, divided by $2^{-i}$, as the separating lines are dropped uniformly at random. Thus we get a bound of $\frac{dr_x^t}{2^{-i}}$ on the probability of recursion terminating.

Alternately, taking a simple union bound, this is at most $t$ times the probability that it happens to any one of the $t$ nearest neighbors. To bound this probability, notice that it is at most the sum of the probabilities that it happens in each of the dimensions. In each dimension, the probability that $x$ is separated from $y$ is simply the absolute value of the length of $x-y$ in that dimension, divided by $2^{-i}$, as the separating lines are dropped uniformly at random. Summing these absolute values gives us $\|x-y\|_1/2^{-i} \leq d^{1/2}\|x-y\|/2^{-i}$, which for each of the $t$-neighbors is at most $d^{1/2} r_x^t/2^{-i}$. Thus the probability that the $t$ nearest neighbors of $x$ do not land in the same level $i$ cell is also bounded by $td^{1/2} r_x^t/2^{-i}$.

If this event occurs, the contribution to $\Delta_x$ would be $d^{1/2} 2^{-i}$, and so the expected contribution is $\min\{d^{\frac{3}{2}}, td\} r_x^t$. There are at most $\log(1/r_x^t)$ levels we must worry about (after which, $\Delta_x < r_x^t$), yielding the stated bound. □

## 3.2 Randomized Histograms for Round Distributions

Recall that at any level in the recursion, we pick $m = 4d \cdot 8^d$ points $p_1, p_2, \ldots, p_m$ uniformly at random from the cell $C$. The clusters are the Voronoi cells defined by these centers, i.e. cluster $C_i$ consists of all points $x \in C$ such that $\|x - p_i\| \leq \|x - p_j\|$ for all $j \neq i$ (breaking ties arbitrarily). In Section 2.3 we showed that these cells are well-rounded.

We now show that for any point $x$ and any $r$, $B(x,r)$ is cut with probability proportional to $r$. The main idea behind the proof is that if the distance $r$ is much smaller than the distance $\ell$ from $x$ to its closest center, then any point $y \in B(x,r)$ gets assigned to a different center only if this center falls within a thin shell of thickness $r$ around $B(x,\ell)$; the probability of this event is proportional to $r$.

Formally, given a clustering $C_1, \ldots, C_k$ and a set $S \subseteq \Re^d$, we say that $S$ is *cut* by the clustering if there are distinct indices $i$ and $j$ such that $S \cap C_i$ and $S \cap C_j$ are both non-empty. We get:

**Lemma 3.2.** *For any convex set $C$ of radius $\rho$ with $x \in C$, $\forall r$, the probability that $B(x,r)$ is cut by a Voronoi partition of $C \in O(dr/\rho)$.*

The lemma states that at any level $i$, the probability that a ball of radius $r_x^t$ around $x$ is cut is proportional to $\frac{dr_x^t}{2^{-i}}$. When this happens, $x$ lies in a cell of diameter about $2^{-i}$. Then by the argument in the proof of Theorem 3.1, the contribution of this level to the expected cell diameter is $O(dr_x^t)$. Combining this with the size of the final cell, we get the following lemma.

**Lemma 3.3.** *Suppose that the number of recursive steps in the construction of the histogram is at most $k$. Then, $\forall x$, $E[\Delta_x] = O(kdr_x^t + 2^{-k})$.*